# A Survey of Impedance Measurement Methods in Power Electronics

Huamin Jie, Zhenyu Zhao, Fan Fei, Kye Yak See
*School of Electrical and Electronic Engineering*
*Nanyang Technological University*
Singapore
Jieh0002@e.ntu.edu.sg

Rejeki Simanjorang, Firman Sasongko
*Rolls-Royce Singapore Pte Ltd*
Singapore

*Abstract*—Impedance is one of the vital parameters that provides useful information for many power electronics related applications. A lot of impedance measurement methods in power electronics have been reported. However, a comprehensive investigation among these methods in terms of their characteristics, advantages, and limitations has not been found in the literature. In order to bridge this gap, a survey of the impedance measurement methods is conducted in this paper. These methods are introduced, discussed, and then classified into different categories depending on the measurement modes, principles, and instruments. Moreover, recommendations for the future research on the impedance measurement are also presented.

## I. INTRODUCTION

Power electronics serve as a key technology for energy conversion, which has been used in various industrial applications, such as power supplies, converters, battery chargers, and motor drive system [1]. As an essential parameter, impedance provides useful information for many power electronics related applications. For example, the impedance of a switched-mode power supply (SMPS) contributes to its systematic electromagnetic interference (EMI) filter design [2]. The impedance of a power grid supports the control operation decision and the stability construction of a grid-connected power converter [3]. The impedance of a power semiconductor device helps to extract its stray inductances and parasitic capacitances for its switching behavior estimation [4], [5]. The impedance of an inverter-fed induction motor is capable for its stator winding insulation faults detection [6]. In view of the significance and the importance of impedances of power electronics systems and devices, extracting their impedance information is very necessary.

For impedance extraction, there are mainly two categories that have been reported, namely simulation-based methods and measurement-based methods. In the simulation-based methods, numerical models are constructed to estimate the impedance of a power electronics system (or device) [7]-[9]. However, they require detailed geometrical and material information of systems (or devices) for model construction, which might not always be available due to intellectual property protection [10]. In contrast, the measurement-based methods can extract impedances in a rapid and straightforward way, and therefore, many of them have been proposed and well applied in many applications. Nevertheless, a comprehensive review in terms of the characteristics, advantages, and limitations of these measurement-based methods has not been discussed in the literature. To apply these methods effectively, this paper presents a survey of impedance measurement methods in power electronics.

In this survey, impedance measurement methods are classified into offline measurement methods and online measurement methods according to the disparity on the measurement mode. Moreover, in line with their principles and instruments, the offline measurement methods are sub-divided into impedance analyzer (IA)-based approach and vector network analyzer (VNA)-based approach. The online measurement methods are subdivided into voltage-current approach, capacitive coupling approach, and inductive coupling approach. This paper is organized as follows. Section II introduces the offline measurement methods. Section III elaborates the online measurement methods. Finally, Section IV concludes this paper and present some future research topics on impedance measurement methods in power electronics.

## II. OFFLINE MEASUREMENT METHODS

The offline impedance measurement methods have been well used to extract impedances of many power electronics related passive components [11]. For instance, the impedance of a common-mode (CM) choke in a power converter can be used to construct the behavioral model for EMI simulations [12], [13]. Parasitic parameters of power semiconductor devices can be extracted from impedance measurements for switching characteristics analysis [14]-[16]. The impedance measurement for a motor drive system helps to build its equivalent circuit model [17] and perform faults diagnosis [18]. Based on the differences in principles and instruments, the existing offline impedance measurement methods can be mainly classified into two categories, namely IA-based approach and VNA-based approach, which are elaborated in Subsections A and B, respectively.

### A. Impedance analyzer-based approach

For the IA-based approach, it can be implemented by either the auto-balancing bridge technique [19] or the radio frequency (RF) voltage-current (V-I) technique [20]. The auto-balancing bridge is developed from the Wheatstone bridge. Fig. 1(a) shows the Wheatstone bridge which contains an oscillator (OSC), a detector (D), and three known resistors ($R_1$, $R_2$ and $R_3$) together with an unknown resistor under test ($R_X$) [21]. When the reading of the detector shows zero, it indicates that no current flows through the detector, and $R_X$ can be calculated by:

$$R_X = \frac{R_2 R_3}{R_1} \qquad (1)$$

Based on the Wheatstone bridge, the auto-balancing bridge is developed to improve the impedance measurement range and realize the automatic control [22]-[24]. Fig. 1(b) shows a schematic diagram of the auto-balancing bridge [25]. A reference resistor $R_{\text{ref}}$ is adopted as the mirror of the system under test (SUT). To maintain the potential of operational amplifier (G) inverting pin (a) to zero, the current flow through the SUT ($I_1$) balances with the current flow through $R_{\text{ref}}$ ($I_2$). By using the measured voltages at the high terminal ($V_1$) and at the output of G ($V_2$), the impedance of the SUT can be calculated by:

$$Z_{\text{SUT}} = -\frac{V_1}{V_2} R_{\text{ref}} \quad (2)$$

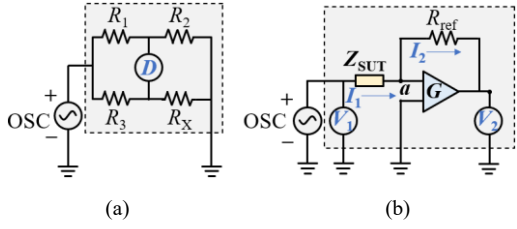

Fig. 1. Schematic digrams. (a) Wheatstone bridge. (b) Auto-balancing bridge.

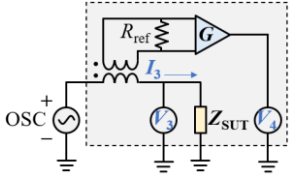

Fig. 2. Schematic digram of RF V-I technique.

In practical applications, the auto-balancing bridge configuration has been further improved to expand the measurement frequency range by containing more accessories. For example, the auto-balancing bridge in the Keysight IA (E4990A) [19] contains many sophisticated devices like null detector, phase detector, and vector modulator to make it as a good candidate in impedance measurements with frequency range of 20 Hz to 120 MHz.

In some cases, the impedance information at higher frequencies (i.e. > 120 MHz) is necessary [26]. Fig. 2 shows the schematic diagram of the RF V-I technique. The RF V-I technique supports the IA to measure impedances at a higher frequency range (e.g. 1 MHz-3 GHz) by measuring the voltage across the SUT ($V_3$) and the current flowing through it ($I_3$), and then combining with Ohm's Law [27], [28]. A vector voltage meter is used to extract $V_3$, and another vector voltage meter together with a known precision reference resistor $R_{\text{ref}}$ and a balun transformer are employed for $I_3$ extraction. By measuring the voltage at the output of the operational amplifier ($V_4$), $I_3$ can be extracted by [27]:

$$I_3 = \frac{V_4}{R_{\text{ref}}} \quad (3)$$

Based on Ohm's Law and extracted $V_3$ and $I_3$, $Z_{\text{SUT}}$ can finally be calculated by:

$$Z_{\text{SUT}} = \frac{V_3}{I_3} \quad (4)$$

It should be noted that the RF V-I technique is usually effective for frequencies higher than 1 MHz due to the intrinsic characteristics of the balun transformer [27].

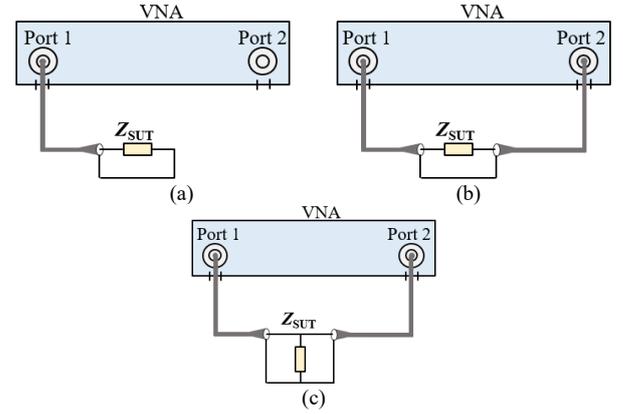

Fig. 3. Schematic diagrams of VNA-based approach. (a) Reflection method. (b) Series-through method. (c) Shunt-through method.

*B. Vector network analyzer-based approach*

The VNA-based approach is applied to extract impedances of SUTs by using the measured scattering parameters (*S*-parameters) [29]-[31]. Based on the differences in principles, the VNA-based approach can be subdivided into three categories: the reflection method, the series-through method, and the shunt-through method. Fig. 3 shows the schematic diagrams of these methods.

The reflection method shown in Fig. 3(a) regards a system as a one-port network that uses reflection coefficient $S_{11}$ to calculate the impedance of the SUT by:

$$Z_{\text{SUT}} = Z_0 \frac{1 + S_{11}}{1 - S_{11}} \quad (5)$$

where $Z_0$ is the characteristic impedance (usually 50 Ω). This method can support a frequency range up to 110 GHz with a relatively good measurement accuracy when the impedance of the SUT is close to 50 Ω [32], [33].

For those SUTs whose impedances are much higher than 50 Ω, the series-through method shown in Fig. 3(b) is preferred. In this method, $Z_{\text{SUT}}$ can be calculated by:

$$Z_{\text{SUT}} = 2Z_0 \left(\frac{1}{S_{21}} - 1\right) \quad (6)$$

In contrast, when the SUT impedance is much lower than 50 Ω, the shunt-through method shown in Fig. 3(c) fits the low-impedance measurements. The calculation formula is given by:

$$Z_{\text{SUT}} = Z_0 \frac{S_{21}}{2(1 - S_{21})} \quad (7)$$

Besides, the calibration procedure also acts as an important role for the impedance measurement accuracies [34], [35]. So far, the most prevalent procedure is the Open-Short-Load (OSL) calibaration, which has been detailed in references [36], [37] and will not be repeated here.

### III. ONLINE MEASUREMENT METHODS

In addition to the offline measurement methods, the online measurement methods extract the impedance of an energized SUT, which can be used for evaluating the actual operating conditions and characteristics of the SUT. For example, the online impedance of a SMPS provides the useful information for its systematic EMI filter design [2]. The online impedance of the stator winding in a motor drive system helps to detect its insulation faults [6]. The online impedance of a grid-connected power converter is useful for making proper control decisions [38]. For online impedance measurement, a lot of methods have been well established, which can be generalized into three categories [39], including voltage-current (V-I) approach, capacitive coupling approach, and inductive coupling approach.

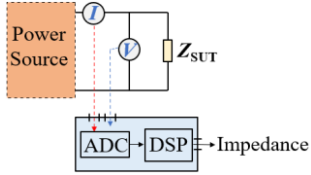

Fig. 4. Schematic diagram of V-I approach.

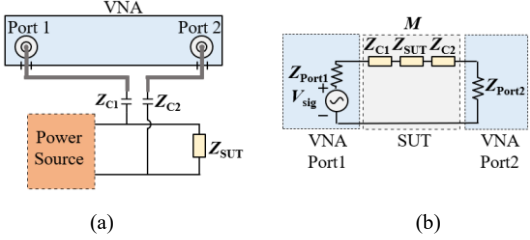

Fig. 5. VNA-based capacitive coupling approach. (a) Measurement setup. (b) Equivalent circuit.

*A. Voltage-current approach*

Fig. 4 shows a schematic digram of the V-I approach. The V-I approach obtains the online impedance of an energized SUT by using a voltage sensor to extract the test signal voltage across the SUT and applying a current sensor to extract the current flowing through it [40]. The information of the voltage and current are extracted by an analog-to-digital converter (ADC) and a digital signal processor (DSP). Eventually, the online impedance is determined with Ohm's Law.

Based on differences of the test signals, the V-I approach can be further divided into passive methods and active methods. The passive methods employ the existing harmonics presented in the energized SUT as the test signal. For the cases with significant background noise, the passive methods may suffer from a low signal-to-noise ratio (SNR) and hence could affect the impedance measurement accuracy [41]. Besides, these methods can only measure impedances at harmonic frequencies. In contrast, the active methods perform online impedance measurement by injecting a transient or steady-state test signal with a specific signal generation device [42]. By selecting the level and spectrum of the test signal properly, the online impedance can be measured at frequencies of interest accurately.

For the V-I approach, various digital signal processing algorithms have been developed, among which discrete Fourier transform (DFT) is widely used [43]. However, the interharmonic distortions and transient events will affect the performance of the DFT due to spectral leakage and picket fence effects. To overcome these limitations, continuous wavelet transform (CWT) with an adjustable window has been proposed [44]. Due to its ability of controlling the selection of scale ranges, CWT is useful for identifying the resonance peak points in the SUT impedance. Nevertheless, CWT provides high redundancy of information for real-time applications and its calculation is complex. In addition, a discrete wavelet packet transform (SDWPT) together with its improved forms have also been used in this approach [41], [42].

*B. Capacitive coupling approach*

The capacitive coupling approach extracts the online impedance of an energized SUT by using coupling capacitors that connects with an IA [45] or a VNA [46]. Fig. 5(a) shows a measurement setup of the VNA-based approach. The coupling capacitors provide a low impedance path for the RF test signal, which is generated by measurement apparatus. On the contrary, the coupling capacitors present high impedance characteristics at low frequencies (LF), so they act as a block to prevent direct current (DC) or LF alternating current (AC) power into the measurement apparatus [47].

Based on the two-port network analysis, Fig. 5(b) shows the equivalent circuit of Fig. 5(a), in which $V_{sig}$ represents the source voltage produced by the VNA; $Z_{Port1}$ and $Z_{Port2}$ are internal impedances of the VNA's port 1 and port 2, respectively, whose values are usually 50 $\Omega$; $Z_{C1}$ and $Z_{C2}$ are the impedances of the coupling capacitors, which can be pre-determined prior to the measurement; $M$ is a two-port network including the coupling capacitors and the SUT. By expressing $M$ with transmission parameters (*ABCD* parameters), it can be obtained by:

$$M = \begin{bmatrix} A & B \\ C & D \end{bmatrix} \quad (8)$$

Based on the network analysis theory [48], $B$ in (8) can be further expressed as:

$$B = Z_{SUT} + Z_{C1} + Z_{C2} \quad (9)$$

Parameter $B$ can be directly obtained via the measured *S*-parameters by the VNA as shown below:

$$B = 50 \cdot \frac{(1+S_{11})(1+S_{22}) - S_{12}S_{21}}{2S_{21}} \quad (10)$$

According to (9) and (10), together with known $Z_{C1}$ and $Z_{C2}$, $Z_{SUT}$ can finally be determined based on the measured *S*-parameters as:

$$Z_{SUT} = 50 \cdot \frac{(1+S_{11})(1+S_{22}) - S_{12}S_{21}}{2S_{21}} \\ -(Z_{C1} + Z_{C2}) \quad (11)$$

For the IA-based capacitive coupling approach, instead of measuring the *S*-parameters, the resultant series impedance of $Z_{SUT}$, $Z_{C1}$, and $Z_{C2}$ can be measured by the IA directly. By de-embedding $Z_{C1}$ and $Z_{C2}$ from the measured resultant series impedance, $Z_{SUT}$ can be obtained correspondingly [49].

*C. Inductive coupling approach*

In contrast to the above-mentioned two approaches, the inductive coupling approach does not require any physical electrical contact with the energized SUT during online impedance measurements, and hence, it eliminates potential electrical safety hazards and simplifies the onsite implementation [10]. The inductive coupling approach was first proposed for power line online impedance measurement [50] and then it has been applied for many other applications, such as EMI filter design for a SMPS [2], condition monitoring of a transformer [51], faults diagnosis of a motor drive system [6], and radiated EMI estimation of a photovoltaic (PV) system [52]. Based on the differences in applications, the measurement setups of this approach can be classified into three categories: single-probe setup (SPS), two-probe setup (TPS), and multi-probe setup (MPS).

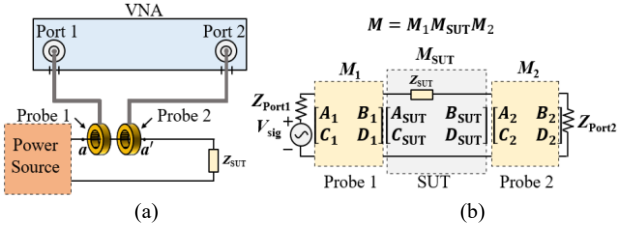

Fig. 6. Frequency-domain TPS. (a) Schematic diagram. (b) Equivalent circuit based on the cascaded two-port network theory.

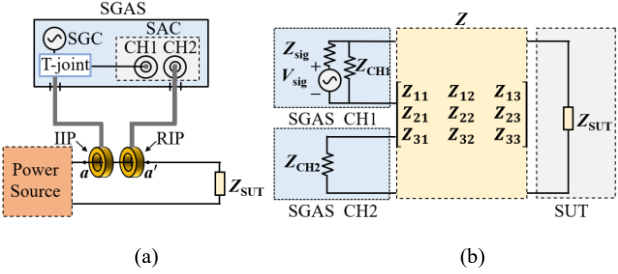

Fig. 7. Time-domain TPS. (a) Schematic diagram. (b) Equivalent circuit based on three-port network theory.

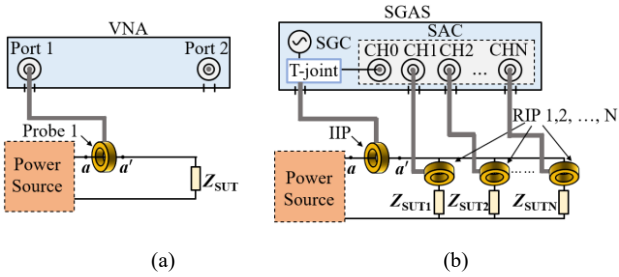

Fig. 8. Schematic diagram. (a) Frequency-domain SPS. (b) Time-domain MPS.

The TPS typically includes two clamped-on inductive probes and a frequency-domain or time-domain measurement instrument. Fig. 6(a) shows a schematic diagram of a frequency-domain TPS, where the VNA is selected as the frequency-domain measurement instrument. One port of the VNA produces a swept-sine excitation signal and then the signal is injected into the energized SUT through an inductive probe. Another port of the VNA monitors the response of the same signal through another inductive probe. By establishing the relationship between the excitation and the response signals, the online impedance of the SUT can be determined.

Fig. 6(b) shows the equivalent circuit model of Fig. 6(a) based on the cascaded two-port network theory [48]. $M_1$ and $M_2$ represent the two-port networks of the probe 1 and probe 2 with the respective clamped wire. $M_{SUT}$ is the two-port network of the SUT. All the $M_1$, $M_2$, and $M_{SUT}$ are expressed in terms of ABCD parameters, and $M$ represents the resultant two-port network of $M_1$, $M_2$, and $M_{SUT}$. Among the two-port networks, $M_1$ and $M_2$ can be pre-characterized by a specific test fixture prior to the online measurement [53]. According to reference [48], $M_{SUT}$ can be rewritten as:

$$M_{SUT} = \begin{bmatrix} A_{SUT} & B_{SUT} \\ C_{SUT} & D_{SUT} \end{bmatrix} = \begin{bmatrix} 1 & Z_{SUT} \\ 0 & 1 \end{bmatrix} \quad (12)$$

$M$ can be directly obtained using the VNA through converting the measured S-parameters into ABCD parameters. According to the known $M_1$ and $M_2$, together with the online measured $M$, $M_{SUT}$ can be obtained correspondingly. Furthermore, based on (12) $Z_{SUT}$ can finally be determined by:

$$\begin{aligned}&Z_{SUT}\\&= \frac{-B_2 D_1 A_{SUT} + B_1 B_2 C_{SUT} + D_1 A_2 B_{SUT} - B_1 A_2 D_{SUT}}{(A_1 D_1 - B_1 C_1)(A_2 D_2 - B_2 C_2)}\end{aligned} \quad (13)$$

Fig. 7(a) shows a schematic diagram of the time-domain TPS, which basically consists of an injecting inductive probe (IIP), a receiving inductive probe (RIP), and a computer-controlled signal generation and acquisition system (SGAS) [54].

To measure $Z_{SUT}$, a single-sine excitation signal produced by the signal generation card (SGC) of SGAS is injected into the SUT through IIP, and the response of the same signal is monitored by RIP. Channel 1 (CH1) and channel 2 (CH2) of the signal acquisition card (SAC) of SGAS are used to sample the time-domain voltages at the input of the IIP ($V_{CH1}$) and the output of the RIP ($V_{CH2}$), respectively. Unlike the frequency-domain TPS that uses the cascaded two-port network theory, a three-port network theory is used for the time-domain TPS to evaluate the influence of the probe-to-probe coupling from the two inductive probes [55]. The probe-to-probe coupling could result in potential effects on the measurement accuracy of the TPS when the two inductive probes are very close to each other due to space constraints. Fig. 7(b) shows the equivalent circuit of Fig. 7(a). $Z$ is a three-port network that takes into account all effects (e.g. probe-to-wire coupling and probe-to-probe coupling) of the IIP and RIP as well as their wires being clamped, which is represented by impedance parameters (Z-parameters). Based on the equivalent circuit, $Z_{SUT}$ can finally be expressed as a function of $V_{CH1}$ and $V_{CH2}$ as follows:

$$Z_{SUT} = \frac{a_3 \left(\frac{V_{CH1}}{V_{CH2}}\right) - a_2}{-\frac{V_{CH1}}{V_{CH2}} + a_1} \quad (14)$$

where $a_1$, $a_2$, and $a_3$ are the TPS frequency-dependent characteristic parameters, which are expressed as:

$$a_1 = \frac{Z_{11}}{Z_{21}} \cdot \left(1 - \frac{Z_{22}}{Z_{CH2}}\right) + \frac{Z_{12}}{Z_{CH2}} \quad (15)$$

$$\begin{aligned}a_2 =& \left(Z_{13} - \frac{Z_{11} Z_{23}}{Z_{21}}\right) \cdot \left[\frac{Z_{31}}{Z_{21}} \cdot \left(1 - \frac{Z_{22}}{Z_{CH2}}\right) + \frac{Z_{32}}{Z_{CH2}}\right]\\&- \left(Z_{33} - \frac{Z_{31} Z_{23}}{Z_{21}}\right) \cdot \left[\frac{Z_{11}}{Z_{21}} \cdot \left(1 - \frac{Z_{22}}{Z_{CH2}}\right) + \frac{Z_{12}}{Z_{CH2}}\right]\end{aligned} \quad (16)$$

$$a_3 = \frac{Z_{31} Z_{23}}{Z_{21}} - Z_{33} \quad (17)$$

where $Z_{CH1}$ and $Z_{CH2}$ represent the internal impedances of CH1 and CH2 of the SAC, respectively.

From (14), $Z_{SUT}$ can be determined by the measured $V_{CH1}$ and $V_{CH2}$ once $a_1$, $a_2$, and $a_3$ are known. In observation of (15)-(17), the frequency-dependent $a_1$, $a_2$, and $a_3$ are determined by the impedance parameters of the three-port network $Z$ and $Z_{CHi}$ ($i = 1, 2$), which keep unchanged for a given TPS. To characterize $a_1$, $a_2$, and $a_3$ of a specific TPS, some calibration techniques have been proposed [39], [55].

Although calibration techniques for the TPS have been proposed to de-embed the influence of the probe-to-probe coupling on the measurement accuracy, the probe-to-probe coupling still exists. To eliminate this coupling fundamentally, an SPS with a frequency-domain measurement has recently been proposed [56]. Fig. 8(a) shows a schematic diagram of the frequency-domain SPS.

By regarding the inductive probe with the wire being clamped as the two-port network, $Z_{SUT}$ can be determined by the measured reflection coefficient using the VNA. The SPS has successfully been used to extract the online noise source impedance of a motor drive system [57], [58].

References [59] and [60] proposed the MPS for simultaneous online impedance measurement of multiple SUTs in multi-branches powered by the same power source. In reference [59], a frequency-domain MPS is proposed that consists of three inductive probes and a VNA, which can only measure the online impedances of two SUTs simultaneously. Moreover, it ignores the parasitic parameters of the inductive probes, which may affect the measurement accuracy. To overcome these issues, reference [60] proposed a time-domain MPS, which consists of one IIP, multiple RIPs depending on the number of SUTs to be measured, and a computer-controlled SGAS, as shown in Fig. 8(b). The time-domain MPS enables measurement of the online impedances of multiple SUTs in multi-branches simultaneously even when the number of branches is larger than two.

It should be noted that for online impedance measurement of an energized SUT with strong background noise and surges, the above-mentioned measurement setups of the inductive coupling approach can incorporate signal amplification and surge protection devices to improve its SNR and ruggedness. However, the principles of the measurement setups still remain the same [6], [56].

IV. DISCUSSIONS, CONCLUSIONS AND FUTURE WORKS

*A. Discussions and conlusions*

This paper has presented a comprehensive survey of various impedance measurement methods in power electronics. Among the offline impedance measurement methods, the auto-balancing bridge technique can perform measurement from several milliohm to a few tens megaohm in a frequency range from a few tens hertz (e.g. 20Hz) to around one hundred MHz [19]. Compared with the auto-balancing bridge technique, the RF V-I technique has a higher end of frequency range but at the expense of the low end of frequency range and the high end of impedance range [20]. In contrast with the aforementioned IA-based methods, the VNA-based methods sacrifice a certain measurement accuracy to achieve a wider measurable frequency range. For instance, some types of VNA can support the available measurement up to 110 GHz [61]. In addition, the reflection, series-through, and shunt-through methods can be selected properly for SUTs with different impedance ranges.

Compared with the offline impedance measurement methods, the implementation of the online impedance measurement methods is relatively complicated, but the measured online impedance can reflect the actual operation condition and characteristics of the SUT. All the three prevalent online impedance measurement approaches (i.e. V-I approach, capacitive coupling approach, and inductive coupling approach) have been well established with good measurement accuracy for specific applications. However, the voltage sensor used in the measurement setup of the V-I approach and the coupling capacitors used in the measurement setup of the capacitive coupling approach require the physical electrical contact to the SUT for online impedance measurement, leading to potential electrical safety hazards especially when the SUT is energized by high voltage. In contrast, the measurement setups (SPS, TPS, and MPS) of the inductive coupling approach have no physical electrical contact with the energized SUT and the clamped-on inductive probes of the measurement setups can easily be mounted on or removed from the wiring connection of the energized SUT. Therefore, this approach eliminates the potential electrical safety hazards and simplify the onsite implementation.

*B. Future works*

This subsection lists some future research topics on impedance measurement in power electronics. For offline impedance measurement, a test fixture serves as a key device. However, the test fixtures in the market are only applicable for components with specific structures and sizes [62]. For those power electronics components with irregular sizes like CM chokes, the commercial test fixtures are usually unavailable. Therefore, it is necessary to develop a systematic method for the design of test fixtures for power electronics components. For online impedance measurement, the signal processing algorithms remains a promising research topic. In addition, the development of novel or improved measurement setups like time-domain SPS is also worth exploring.

ACKNOWLEDGMENT

This study is supported under the RIE2020 Industry Alignment Fund-Industry Collaboration Projects (IAF-ICP) Funding Initiative, as well as cash and in-kind contribution from Rolls-Royce Singapore Pte Ltd.